\documentclass[longauth]{aa} % for the long lists of affiliations 
\usepackage{graphicx}
\usepackage{txfonts}

\begin{document}
   \title{Spitzer observations of the asteroid-comet transition object and potential spacecraft target 107P (4015) Wilson-Harrington}
   \author{J. Licandro
	  \inst{1,2}
	  \and
          H. Campins
          \inst{3}
	  \and
	  M. Kelley
	  \inst{4}
	  \and
	  Y. Fern\'andez
	  \inst{3}
	  \and
	  M. Delb\'o
	  \inst{5}
	  \and
	  W. T. Reach
	  \inst{6}
	  \and
	  O. Groussin
	  \inst{7}
	  \and
	  P. L. Lamy
	  \inst{7}
	  \and
	  I. Toth
	  \inst{8}
	  \and
	  M. F.  A'Hearn
	  \inst{9}
	  \and
	  J. M. Bauer
	  \inst{10}
	  \and
	  S. C. Lowry
	  \inst{10,11}
	  \and
	  A. Fitzsimmons
	  \inst{12}
	  \and
	  C. M. Lisse
	  \inst{13}
	  \and
	  K. J. Meech
	  \inst{14}
	  \and
	  J. Pittichov\'a
	  \inst{14, 15}
	  \and
	  C. Snodgrass
	  \inst{16}
	  \and
	  H. A. Weaver
	  \inst{13} }

   \offprints{J. Licandro}

  \institute{Instituto de Astrof\'{\i}sica de Canarias, c/V\'{\i}a L\'actea s/n, 38200 La Laguna, Tenerife, Spain. \\
              \email{jlicandr@iac.es}
              \and 
              Departamento de Astrof\'{\i}sica, Universidad de La Laguna, E-38205 La Laguna, Tenerife, Spain\\
              \and
              Physics Department, University of Central Florida, Orlando, FL, 32816, USA.\\
	     \and
	     Department of Astronomy, University of Maryland, College Park, MD 20742-2421, USA.\\ 
              \and
              UNS, CNRS, Observatoire de la C\^ote d'Azur, Nice, France.\\
              \and
              IPAC, MS 220-6, Caltech, Pasadena, CA, 91125, USA.\\
	      \and
              Laboratoire d'Astrophysique de Marseille, CNRS \& Universit\'e de Provence, 13388 Marseille Cedex 13, France \\
              \and
              Konkoly Obs., P.O. Box 67, H-1525, Hungary.\\
              \and
              Dept. of Astronomy, Univ. of Maryland, College Park, MD, 20742, USA. \\
              \and
              NASA/JPL, 4800 Oak Grove Dr., Pasadena, CA, 91109, USA.\\
             \and
             Centre for Astrophysics and Planetary Science, Univ. of Kent, Canterbury CT2 7NH, UK             
              \and
              Astrophysics Research Centre, QueenÕs Univ. Belfast, Belfast, BT7 1NN, UK.\\
              \and              
              Johns Hopkins Univ. Applied Physics Lab., 11100 Johns Hopkins Rd, Laurel, MD, 20723, USA. \\
              \and
              Inst. for Astronomy, Univ. of Hawaii, 2680 Woodlawn Dr., Honolulu, HI 96822, USA.\\
              \and
              Astronomical Institute, Slovak Academy of Sciences, Slovak Republic, 845 04 Bratislava.\\
              \and
              ESO, Alonso de C\'ordova 3107, Vitacura, Santiago, Chile.\\
 }
   \date{Received April 2009; accepted}

% \abstract{}{}{}{}{} 
% 5 {} token are mandatory
 
  \abstract
   {Near-Earth asteroid-comet transition object 107P/ (4015) Wilson-Harrington  is a possible target of the  joint European Space Agency (ESA) and Japanese Aerospace Exploration Agency (JAXA)  Marco Polo sample return mission.   Physical studies of this object are relevant to this mission, and also to understanding its asteroidal or cometary nature.}
   {Our aim is to obtain significant new constraints on the surface thermal properties of this object.}
   {We present mid-infrared photometry in two filters (16 and 22 $\mu$m) obtained with NASA's Spitzer Space Telescope on February 12, 2007, and results from the application of the Near Earth Asteroid Thermal Model (NEATM). We obtained high S/N in two mid-IR bands allowing accurate measurements of its thermal emission. }
   {We obtain a well constrained beaming parameter ($\eta$ = 1.39$ \pm$ 0.26) and obtain a diameter and geometric albedo of $D$ = 3.46 $\pm$ 0.32 km, and $p_V$ = 0.059 $\pm$ 0.011. We also obtain similar results when we apply this best-fitting thermal model to single-band mid-IR photometry reported by Campins et al. (1995),  Kraemer et al. (2005) and Reach et al. (2007). 
}
   {The albedo of 4015 Wilson-Harrington is low, consistent with those of comet nuclei and primitive C-, P-, D-type asteorids. 
   %The value of its beaming parameter is high, but compatible with roughly 1/3 of the accurate $\eta$ values determined for several nuclei by Fernandez et al. (2009, in prep.). 
   We establish a rough lower limit for the thermal inertia of W-H of 60 $Jm^{-2}s^{-0.5}K^{-1}$ when it is at $r$=1AU, which is slightly over the limit of  30 $Jm^{-2}s^{-0.5}K^{-1}$  derived by Groussin et al. (2009) for the thermal inertia of the nucleus of comet 22P/Kopff.  
}
   
   \titlerunning{Spitzer Observations of 4015 Wilson-Harrington}
   \maketitle
   
%________________________________________________________________

\section{Introduction}
The asteroid-comet transition object 107P/ (4015) Wilson-Harrington (hereafter W-H) is a possible target of the joint European Space Agency (ESA) and Japanese Aerospace Exploration Agency (JAXA)  Marco Polo sample return mission. This mission seeks to understand the origin and nature of volatile and organic material in the early Solar System and ideally would do this by returning samples of a primitive asteroid -- i.e. a C, P or D asteroid in the Tholen classification system (Tholen and Barucci \cite{Tholen1989}). C, P and D objects are abundant in the middle to outer Main Belt; and are believed to be primitive remnants from the early Solar System.  

W-H was discovered with a cometary appearance  in 1949  (Cunningham \cite{Cunningham50}, Fernandez et al. \cite{Fernandezetal97}). Recovered in 1979, and observed at every apparition since, it has always appeared as a point source. W-H may have a cometary origin, but its cometary nature is not firmly established. Its orbit has an unlikely cometary origin, its Tisserand parameter is 3.084 and it has only a 4\% chance of coming from the Jupiter-family comet population (Bottke et al. \cite{Bottkeetal02}). However, the same argument against a cometary origin could be made for the active comet 2P/Encke, which could have achieved its orbit either due to the influence of nongravitational forces (Fernandez et al. \cite{Fernandez2002}) or gravitational perturbations (Levison et al. \cite{Levisonetal2006}).  Both photometric and spectroscopic data suggest classification of W-H within the C complex (Tholen and Barucci \cite{Tholen1989}, Chamberlin et al. \cite{Chamberetal96}).

Licandro et al. (\cite{Licandroetal07}) suggested a link between W-H, (3200) Phaethon (another similar asteroid-comet transition object in the NEA population), and the objects with asteroidal orbits in the main belt that show cometary-like activity (the ``Main Belt Comets", MBCs, Hsieh \& Jewitt \cite{Hsieh06}), as their colors and/or spectra correspond to those of Cb- or B-type asteroids. Licandro et al. show that Phaethon's surface composition suggests it is unlikely that this object has a cometary origin.

%Another well known NEA in an unlikely cometary orbit, (3200) Phaethon, presented past activity. Phaethon is the parent of the Geminid meteor shower, and, although it was never observed to be active, past activity is suggested by the way it produced the Geminids (Gustafson \cite{Gusta89}; Williams \& Wu \cite{WillWu93}). Licandro et al. (\cite{Licandroetal07}), analyzing the 0.35-2.4$\mu$m spectrum,  provided compositional arguments that favor a main-belt and not a cometary origin for Phaethon. The visible spectrum of W-H obtained by Chamberlin et al. (\cite{Chamberetal96}) is compatible with a Cb- or B-type taxonomy class, the same as Phaethon. 

%Hsieh \& Jewitt (\cite{Hsieh06}) reported that three objects with orbits within or similar to the Themis family have displayed faint cometary activity as well, and called them ``Main Belt Comets" (MBCs). A cometary origin of the MBCs is  dynamically unlikely. Also visible spectra of MBCs (7968) Elst-Pizarro and (118401) LINEAR support that they are Themis family asteroids instead of captured comets, they are Cb or B-type like most of the Themis familiy asteroids (Licandro et al. \cite{Licandroetal07a}) suggesting that they likely have an asteroidal nature.  Licandro et al. (\cite{Licandroetal07}) suggested that Phaethon and W-H may be MBCs dynamically scattered into the NEA population.

If the mechanism that activates the MBCs is driven by the sublimation of ices, that means some asteroids in the outer main-belt could have retained ices below their surface and, under certain conditions, become ``activated asteroids''. This supports the scenario that the {\em snow line} fell somewhere in the asteroid belt around 3AU.  This suggest that asteroids could have contributed to the formation of the oceans on Earth due to impacts on our planet (e.g. Drake and Campins \cite{Drake06},  Mottl et al. \cite{Mottl}). While W-H has shown cometary activity, the existence of MBCs shows that this does not preclude it from having an asteroidal rather than cometary dynamical origin.

Mid-infrared observations are needed in order to derive an accurate size and albedo using the near-Earth asteroid thermal model (NEATM). The surface temperature can be used to constrain the thermal inertia . Information about the surface temperature can be obtained from multi-wavelength filter photometry. 
We present new high S/N mid-infrared broadband photometry of W-H which places significant new constraints on its surface properties, and we compare these results with previously published mid-infrared data.

\section{Observations and data reduction}

W-H was observed as part of the SEPPCoN program (Survey of Ensemble Physical Properties of Cometary Nuclei, see Fernandez et al., \cite{Fernandezetal08}), with the Spitzer Space Telescope  on February 12, 2007. Observations started at 07:36:55 UT and lasted for 17.14 minutes. We used  the peak-up imaging mode
of the InfraRed Spectrograph (IRS) and obtained images in both the "blue" and "red" passbands (monochromatic wavelengths of 15.769 and 22.327 $\mu$m, respectively). The total integration time was 94 s for the "blue" and 147 s for the "red" images. The data were processed and photometrically calibrated with the Spitzer Science Center's IRS pipeline (version 16.1.0). MOPEX software (version 16.2.8) (Makovoz and Khan, \cite{Makovoz05}) was used to clean, register, and co-add the data. Each exposure was aligned onto a common coordinate system, taking into account the motion of W-H, and any constant offsets in the background were removed. The exposures were averaged together, filtering out bad pixels, which produced two final images: one each for the blue and red peak-up arrays. Our data reduction procedure is identical to that followed in our previously described work (Groussin et al. \cite{Groussinetal09}). The object appeared as a point-source in both images. The photometric analysis was then performed by means of point spread function fitting . Color correction to the photometry was applied based on W-H's color temperature. The observing geometry and measured fluxes are in Table \ref{Table1}, where $r$ and $\rho$ are the heliocentric and Spitzer distance respectively, $\alpha$ is the phase angle, $F_{15.8}$ and $F_{22.3}$ are the fluxes measured in the ``blue" and ``red" bands.

\begin{table}
\centering
\begin{tabular}{l c c c c c}
\hline
\hline
Date (UT) & $r$ (AU) & $\rho$ (AU)  & $\alpha$($^{\circ}$) & $F_{15.8} (mJy)$ & $F_{22.3} (mJy)$ \\ \hline
02/12/07 & 4.078 & 3.520 & 12.7 & 1.26$\pm$0.08 & 2.20$\pm$0.04 \\  \hline
\end{tabular}
  \caption{Observing geometry and measured fluxes .}
  \label{Table1}
\end{table}

\section{Thermal modeling}

\begin{figure}
	\centering
	\includegraphics[width=8cm]{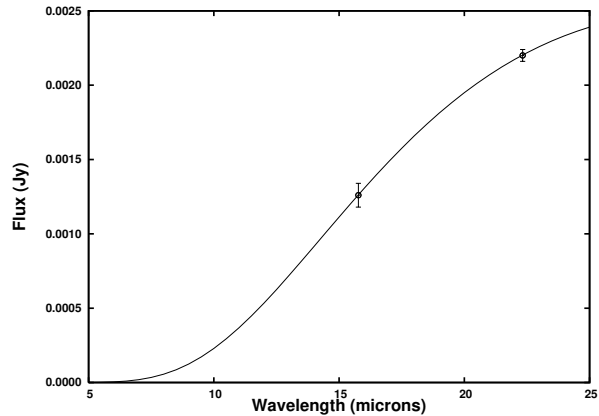}
	\caption{Thermal model fit (solid line) using NEATM to the observed flux (circles) of 4015 Wilson-Harrington.}
 	\label{Fig1}
 \end{figure}

The flux of asteroids in this spectral range is dominated by thermal emission. The measured spectral energy distribution (SED) depends on the object's size, composition, and temperature distribution.  This last term is dependent on several factors, including distance from the Sun, albedo, thermal inertia, surface roughness, shape, rotation rate, and spin-pole orientation.  Many of these are unknown for W-H, therefore fitting our photometry with a thermophysical  model (Muller \cite{MullerPhd} and references therein) is unwarranted. We chose to use the semi-empirical Near-Earth Asteroid Thermal Model (NEATM, Harris \cite{Harris1998})  to fit our 16 and 22 $\mu$m photometric data points in order to derive the size and albedo of the object. We also use the NEATM to derive information on the thermal intertia of W-H.
%NEATM is a refinement of the standard thermal model (STM, Lebofsky et al. \cite{Lebofskyetal1986}; Lebofsky and Spencer \cite{Lebofskyetal1989}), which was developed and calibrated for main-belt asteroids.  NEATM accounts for observations at larger phase angles and possible deviations from the STM in the thermal behavior of smaller asteroids.  Unlike STM, NEATM requires observations at multiple wavelengths and uses this information to force the model temperature distribution to be consistent with the apparent color temperature of the asteroid. 
We use the NEATM to solve simultaneously for the beaming parameter ($\eta$) and the diameter ($D$). The beaming parameter was originally introduced in the STM to allow the model temperature distribution to fit the observed enhancement of thermal emission at small solar phase angles due to surface roughness. In the NEATM, $\eta$ can be thought of as a modeling parameter that allows a first-order correction for any effect that influences the observed surface temperature distribution (such as roughness, thermal inertia, and rotation).  At small phase angles ($\alpha < $30$^{\circ}$) values of $\eta <$ 1 indicate that the presence of surface roughness enhances the temperature with respect to that of  a smooth surface, whereas $\eta > 1$ indicates substantial thermal inertia. The values of $\eta$ , $D$ and the geometric albedo ($p_V$) derived for 4015 W-H using the NEATM (see Fig. \ref{Fig1}) are presented in Table \ref{Table2}. Uncertainties in the diameter estimated with thermal modeling usually exceed the formal errors and are typically 10-15\% (e.g., Delb\'o et al. \cite{Delboetal03}). We assumed a thermal infrared emissivity ($\epsilon$) of 0.9. To estimate $p_V$ we adopted $H_V$ = 15.99 $\pm$ 0.10 and $G$ = 0.15 (MPC 17270), where $H_V$ is the visible absolute magnitude and $G$ is the slope parameter. The listed error in the geometric albedo includes the uncertainty in $H_V$ and the diameter uncertainty propagated into the albedo. The error we assume in $H_V$ (0.1 mag) is due to the rotational lightcurve amplitude of 0.2 magnitudes (Osip et al. \cite{Osip95}). 

\begin{table}
\centering
\begin{tabular}{l c c c c}
\hline\hline
Data  & $D$ (km)  & $\eta$   & $p_V$ & $\alpha$($^{\circ}$) \\  \hline
This paper &3.46$\pm$0.32  & 1.39$\pm$0.26 & 0.059$\pm$0.011 &12.7\\  
Kraemer et al.   & 3.3 $\pm$0.18  & 1.22 &0.063$\pm$0.007 & 51.0 \\
%Kraemer et al. $^{(2)}$  &11.8   $\pm$ 1.7   & -- $^{(5)}$    &0.005 $\pm$ 0.041         &51.0\\ 
\hline
\end{tabular}
\caption{NEATM  Best Fit Model of the dat presented in this paper, and the results from Kraemer et al. (2005). }
\label{Table2}
\end{table}

\section{Discussion and conclusions}

\begin{table*}
\centering
\begin{tabular}{c c c  c c c c c c c}
\hline\hline
Data  &$CWL$ &$Flux$ &$\alpha$($^{\circ}$) & $\eta$=1.39 & & $\eta$=1.13  & & $\eta$=1.65 &  \\ 
 &($\mu$m) &(mJy) & $\deg$ &$D$ (km) & $p_V$ & $D$ (km)  & $p_V$  & $D$ (km)  & $p_V$   \\  \hline
Campins et al. & 10.6 &93$\pm$24 &16.0 & 3.51 $\pm$ 0.45  & 0.057 $\pm$ 0.021 & 3.09 $\pm$ 0.39  
& 0.074 $\pm$ 0.027 & 3.93 $\pm$ 0.39  & 0.046 $\pm$ 0.027 \\  
Reach et al. &23.68 &3.8$\pm$0.2&17.6 & 3.38 $\pm$ 0.09 & 0.062 $\pm$ 0.003 & 3.11 $\pm$ 0.08 
& 0.074 $\pm$ 0.004 & 3.64 $\pm$ 0.10 & 0.054 $\pm$ 0.003 \\  
Kraemer et al.  &8.432 &63$\pm$7 &51.0 & 3.50 $\pm$ 0.17  &0.058 $\pm$ 0.007  & 3.07 $\pm$ 0.17  
&0.075 $\pm$ 0.009  & 3.91 $\pm$ 0.22  &0.046 $\pm$ 0.005 \\  \hline
\end{tabular}
\caption{NEATM Best Fit Models to previously published data assuming a fixed value of $\eta=1.39$ obtained in this paper, and this $\eta$ $\pm$ 1-$\sigma$ values in order to derive a possible range. }
\label{Table3}
\end{table*}

Our results can be compared with previous results. Mid-infrared observations of W-H were reported by Campins et al. (\cite{Campins95}), Kraemer et al. (\cite{Kraemer05}), and Reach et al. (\cite{Reach07}). 

Campins et al.  presented simultaneous J, H, K, and N band broad-band photometry and derived two different values of size and albedo, a diameter of 2.60$\pm$0.16 km and a geometric albedo in the J band of 0.10$\pm$0.02 using the STM,  and a diameter of 3.90$\pm$0.25 km and albedo 0.05$\pm$0.01 using the Isothermal Latitude Model (ILM). 
Kraemer et al.  presented broadband photometry of W-H in two filters centered at 8.3 and 14.7 $\mu$m, using the SPIRIT III instrument on board of the Midcourse Space Experiment, and using NEATM  derived $D$, p$_V$ and $\eta$ (their results are presented in the second line of Table \ref{Table2}). 
Reach et al.  presented Spitzer Space Telescope images using the 24 $\mu$m filter, and reported the measured emission flux in this band, but they did not derive either the size or the albedo.

We first compare our results with those of Kraemer et al. since they obtained photometry in two bands, and applied NEATM to those data (see Table \ref{Table2}). While their $D$ and $p_V$ are consistent with ours, their $\eta$ value is smaller.  Their observations were done at a much higher phase angle, so this doesnot fit well with the trend of increasing $\eta$ with increasing solar phase angle that has been observed for about 40 near-Earth asteroids so far (e.g., Delb\'o \cite{Delbo04}, Campins et al. \cite{Campinsetal09}). However according to Kraemer (personal
communication) they did not use the 14.7 $\mu$m flux to derive their reported $D$ and $p_V$ as they concluded that this flux was poorly constrained. So the $D$ and $p_V$ reported by Kraemer et al. were obtained by fitting NEATM to the 8.3 $\mu$m flux alone while using a fixed value of $\eta$. This means that our computed value of $\eta$ represents the first measurement of W-H's beaming parameter.

Finally, we used the NEATM to fit only the Kraemer et al. flux at 8.3 $\mu$m using our computed value of  $\eta$ = 1.39 $\pm$ 0.26. To incorporate the uncertainty in $\eta$, we used 3 different values $\eta$ = 1.39, 1.13, and 1.65 (= 1.39 $\pm 1\sigma$) respectively.  We did the same with the Campins et al. and Reach et al. observations. Notice that in those cases the phase angle is very similar. In the case of Reach et al., we re-analyzed the images and computed the flux, and after applying the color correction we obtained a value of 3.8 mJy at 23.68 $\mu$m. Results are shown in Table \ref{Table3}, where $CWL$ is the central wavelength of the reported fluxes ($Flux$) .

From Table \ref{Table3}, we can conclude that the $D$, $p_V$ and $\eta$ values we determine with the Spitzer data are robust and consistent with previously published observations. In particular, the albedo obtained is consistent with that of cometary nuclei (Fernandez et al. \cite{Fernandez05}) and of primitive asteroids. It is interesting to note that the $\eta$ value we obtained (1.39) is high at a rather small phase angle. Notice also that, in all cases, to have a cometary like albedo ($p_V <$ 0.075 according to Fernandez et al. \cite{Fernandez05}), W-H's $\eta$ should be larger than 1.1.   

The analysis of the SEPCoN sample by Fernandez et al. (2009 in prep.) report that they have 21 cometary nuclei (including W-H) for which the measured beaming parameter  $\eta$ is probably robust. While the 1-$\sigma$ error bar on each value of eta is comparable to what we report here for W-H ($\pm$0.2 to 0.3), the distribution of $\eta$ values is wide, with two-thirds of the group near 1.0 and the other third straddling 1.4. So, the $\eta$  value of W-H fits well in the second group of comets and cannot also be used to discriminate between the asteroidal or cometary nature of this object.

%We also note that $\eta$=1.39 at $\alpha$=12 degrees is larger than the $\eta$ of any of the NEAs observed by Delb\'o \cite{Delbo04}, Delb\'o et al. (\cite{Delboetal07}) and Campins et al. (\cite{Campinsetal09}) at this phase angle. Delb\'o et al. (\cite{Delboetal07}), based on the analysis of the $\eta$ values determined for their sample of NEAs, concluded that the average thermal inertia of NEAs in the km-size range is 200$\pm$70 $Jm^{-2}s^{-0.5}K^{-1}$, larger than the average thermal inertia of large Main Belt asteroids (between 5 and 25 $Jm^{-2}s^{-0.5}K^{-1}$ according to M{\"u}ller \& Lagerros, \cite{Muller98}). Groussin et al. (\cite{Groussinetal09}) suggest that the thermal inertia of cometary nuclei is low ($\Gamma<$ 50  $Jm^{-2}s^{-0.5}K^{-1}$), indicating a surface probably covered by a fine thick regolith (a dust mantle). The high $\eta$ value we derived for W-H suggests a large $\Gamma$ and could favor a non-cometary nature. But $\eta$ depends also on the rotational properties, orientation of the polar axis, and shape of the object, and there is no pole position determination for W-H that could help us to constrain $\Gamma$.

On the other hand, the value of $\eta$ can be used to estimate the value of the thermal inertia from a relation existing between $\eta$ and the thermal parameter $\Theta = \Gamma \sqrt{2\pi P^{-1}} / (\epsilon \sigma T _{SS}^3)$, where $\Gamma$ is the value of the thermal inertia, $P$ is the rotation period, $\epsilon$ the infrared emissivity, $\sigma$ the Stefan Boltzmann constant, and $T_{SS}$ the subsolar temperature. $T_{SS}$ can be derived from the equation: $(1-A)S_\odot r^{-2} = \eta \epsilon \sigma T_{SS}^4$, where A is the bolometric Bond albedo, and $S_\odot$ is the solar constant at 1 AU. The latter can be calculated from the value of $p_V$ in Table \ref{Table2} using the formula $A=p_V\times(0.29 + 0.684 G)$ then $A$=0.23 (Bowell et al., \cite{Bowel89}) and assuming a slope parameter $G$=0.15.

The derived value of $\eta$ of about 1.4 implies a value of $\Theta$ between 1.5 and 3.5 according to Spencer (\cite{Spencer}; Fig 5), depending on the degree of surface roughness (which is unknown for W-H) and assuming observation at a phase angle $\alpha=0^{\circ}$  (a valid comparison considering that our observations were done at a very low phase angle, and thus the possible variation of $\eta$ with $\alpha$ is much lower than the uncertainties) . This corresponds to a range of thermal inertia between 20 and 70 $J m^{-2} s^{-0.5} K^{-1}$ for our observations obtained at $r$=4.078 AU.

The thermal conductivity in the regolith of an airless body is temperature dependent (Keihm, \cite{Keihm}), as is the thermal inertia. This temperature dependence of $\Gamma$ must be taken into account in order to compare our results with those obtained for NEAs (e.g. Delb\'o et al., \cite{Delboetal07}) that are usually observed closer to the Sun (at about 1AU). Under the assumption that heat is transported in the regolith mainly by radiative conduction between grains, the thermal conductivity is proportional to $T_{SS}^3$ (K{\"u}hrt and Giese, \cite{Kuhrt}; Jakosky, \cite{Jakosky}). In this case $\Gamma \propto T_{SS}^{3/2}$ and thus $\Gamma \propto r^{3/4}$. On the basis of this dependence we find that the range of thermal inertia is about 60 and 200 $Jm^{-2}s^{-0.5}K^{-1}$ at $r$=1 AU. These values are between the typical thermal inertias of km-sized NEAs (Delb\'o et al., \cite{Delboetal07}) and those of cometary nuclei (Groussin et al. \cite{Groussinetal09}). We caution here that the values above should be taken as indicative because they assume a spin-pole orientation perpendicular to the orbital plane and $\alpha=0^{\circ}$. Unfortunately the spin-pole of W-H is unknown, so larger values of $\Gamma$ cannot be ruled out. Moreover, $<$ 5\% lower values of $\Gamma$ would have been derived if the morning side (at $\alpha = 12.7^{\circ}$) of the body was actually observed. 

In conclusion, the albedo of 4015 Wilson-Harrington is low ($p_V$=0.059 $\pm$ 0.011), consistent with those of comet nuclei and primitive C-, P-, D-type asteorids. The value of its beaming parameter ($\eta$=1.39$\pm$0.26) is  compatible with about 1/3 of the 21 comet nuclei with accurate $\eta$ values determined by Fernandez et al (2009, in prep.). From $\eta$, we derived a lower limit for the W-H thermal inertia of W-H of 60 $Jm^{-2}s^{-0.5}K^{-1}$, which is slightly over the limit of 30 $Jm^{-2}s^{-0.5}K^{-1}$ derived by Groussin et al. (\cite{Groussinetal09}) for the thermal inertia of the nucleus of comet 22P/Kopff.  

%The orbit, the spectral similarities with the other paradigmatic case of asteroid-comet transition object in the NEA population, (3200) Phaethon, and differences with the published spectra of cometary nuclei, the high beaming parameter, and the rapid rotation period (6.1hr), considered all together, favor an asteroidal nature for Wilson-Harrington.

{\em Acknowledgements:}
We thanks the referee J. Emery for his useful comments, and K. Hargrove for reviewing the manuscript.
JL gratefully acknowledges support from the spanish ``Ministerio de Ciencia e Innovaci\'on'' projects AYA2005-07808-C03-02 and AYA2008-06202-C03-02.
HC gratefully acknowledges support from NASA's Spitzer Science Center, Jet Propulsion Laboratory and Planetary Astronomy program.  HC was a visiting Fulbright Scholar at the ``Instituto de Astrof\'{\i}sica de Canarias'' in Tenerife, Spain. JP gratefully acknowledges support from  the Slovak Academy of Sciences Grant VEGA 2/7040/27.

\end{document}